\documentclass{ws-rv961x669}
\usepackage{ws-rv-van}     % numbered citation/references (default)
\usepackage{ws-rv-thm}     % comment this line when `amsthm / theorem / ntheorem` package is used
\usepackage{subfigure}     % required only when side-by-side / subfigures are used
\newcommand{\APNY}[1]{Ann. Phys. (N.Y.) {\bf {#1}}}

\newcommand{\EPJA}[1]{Eur. Phys. J. A {\bf {#1}}}

\newcommand{\JPG}[1]{J. Phys. {\bf G{#1}}}
\newcommand{\MPLA}[1]{Mod. Phys. Lett. {\bf A{#1}}}
\newcommand{\NPA}[1]{Nucl. Phys. {\bf A{#1}}}
\newcommand{\NPB}[1]{Nucl. Phys. {\bf B{#1}}}
\newcommand{\PLB}[1]{Phys. Lett. {\bf B{#1}}}
\newcommand{\PRev}[1]{Phys. Rev. {\bf {#1}}}
\newcommand{\PRC}[1]{Phys. Rev. C {\bf {#1}}}
\newcommand{\PRD}[1]{Phys. Rev. D {\bf {#1}}}
\newcommand{\PRL}[1]{Phys. Rev. Lett. {\bf {#1}}}

\newcommand{\PPN}[1]{Phys. Part. Nucl. {\bf {#1}}}

\newcommand{\ZP}[1]{Z. Phys. {\bf {#1}}}
\newcommand{\ZPA}[1]{Z. Phys. A, At.\& Nucl. {\bf {#1}}}

\newcommand{\oh}{\frac{1}{2}}
\newcommand{\ot}{\frac{1}{3}}
\newcommand{\beq}{\begin{equation}}
\newcommand{\eeq}{\end{equation}}
\newcommand{\beqa}{\begin{eqnarray}}
\newcommand{\eeqa}{\end{eqnarray}}
\newcommand{\nn}{\nonumber}

\newcommand{\alp}{\alpha}
\newcommand{\lam}{\lambda}

\newcommand{\bd}[1]{ \mbox{\boldmath $#1$}}

\newcommand{\bessk}[2]{K_{#1}\left ( #2\right )}
\newcommand{\bessi}[2]{I_{#1}\left ( #2\right )}

\makeindex

\begin{document}

\chapter[The Fullerene-like Structure of Superheavy Element $Z=120$ (Greinerium)-a tribute to Walter Greiner]
{The Fullerene-like Structure of \\
Superheavy Element $Z=120$ (Greinerium)
\\-a tribute to Walter Greiner \label{ra_ch1}}

\author[\c S. Mi\c sicu]{\c S. Mi\c sicu\footnote{Author footnote.}}

\address{Department for Theoretical Physics\\
National Institute for Nuclear Physics and Engineering,\\
P.O.Box MG-6, RO-077125 Bucharest-Magurele, \\
misicu@theory.nipne.ro\footnote{Affiliation footnote.}}

\author[I. N. Mishustin]{I.N. Mishustin\footnote{Author footnote.}}

\address{Frankfurt Institute for Advanced Studies\\
J. W. von Goethe University,\\
Frankfurt am Main , Germany,\\
\&\\
Russian Research Center, Kurchatov Institute, \\
123182 Moscow, Russia\\
mishustin@fias.uni-frankfurt.de\footnote{Affiliation footnote.}}

\begin{abstract}
We review the basic ideas and some theoretical models  behind the concept of fullerene-like structures made of $\alp$-particles.
The possibility of such a peculiar nuclear shape developing in a double magic superheavy nucleus with $Z$=120
was mentioned for the first time in the literature by Walter Greiner and then constantly defended by him.
We provide estimates of the energy of such metastable states within the liquid-drop model. 
In the second part of this paper we discuss a simple model rooted in the nontopological soliton  
model consisting of a complex scalar field describing a finite system of $\alp$ bosons with non-linear
self-interactions coupled to the electromagnetic field. 
We demonstrate that this model predicts density depletion in the central region of a soliton-like structure 
for a range of model parameters. 
\end{abstract}

\body

\section{Introduction}\label{ra_sec1}

More than two decades ago, Walter Greiner conjectured that 
some superheavy elements assume a fullerene-like structure formed of 
$\alpha$-clusters. This speculation was born out from relativistic
mean-field (RMF) calculations for nuclei with charges around $Z=$120
which are pointing to a pronounced central density depletion 
\cite{bend99}. In the following years he constantly advocated this exotic picture
in superheavies \cite{gre01,mis02}.

\begin{figure}[h]
\centerline{
\subfigure[]{\includegraphics[width=5.6cm]{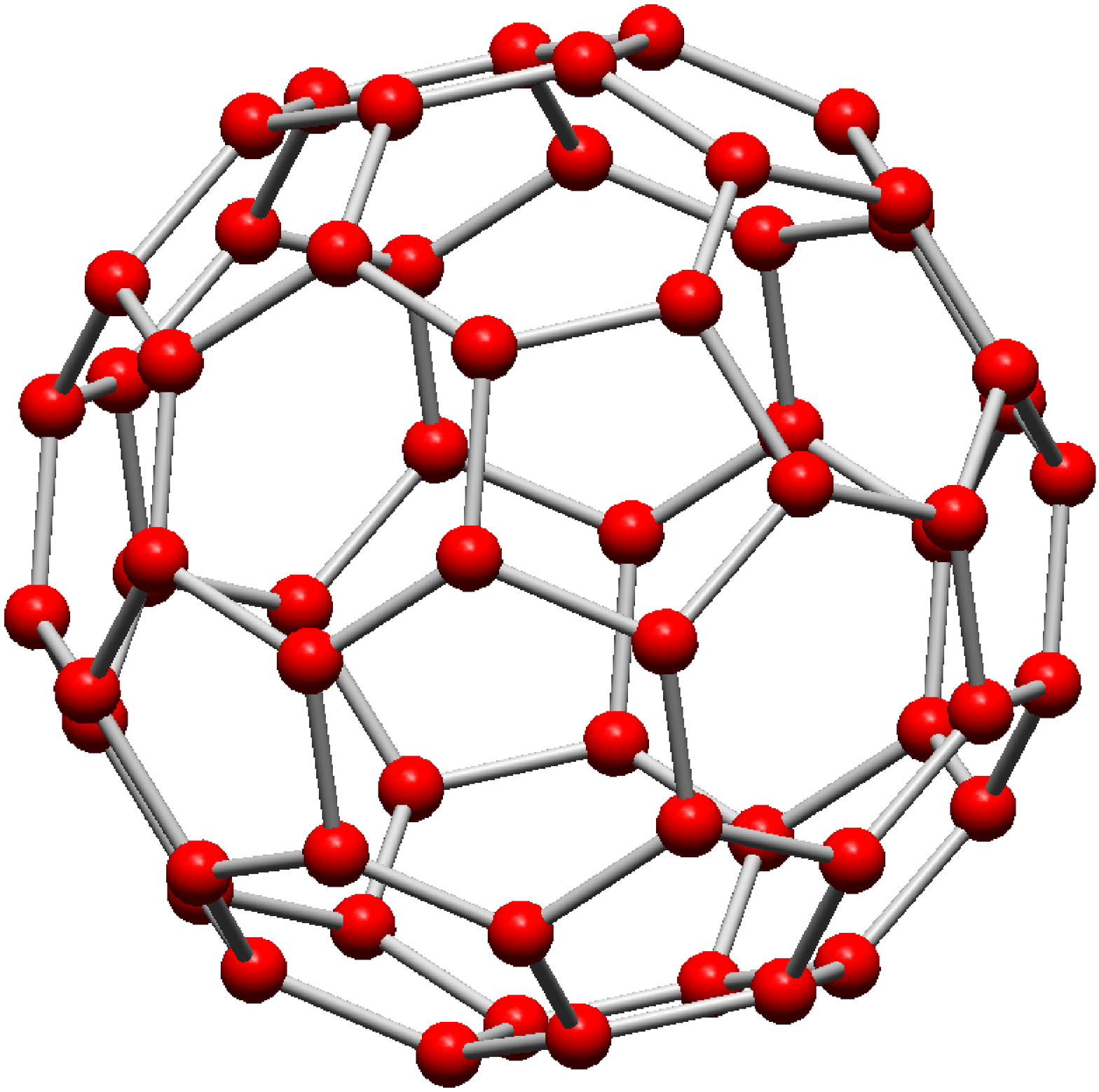}}
\hspace*{25pt}
\subfigure[]{\includegraphics[width=5.6cm]{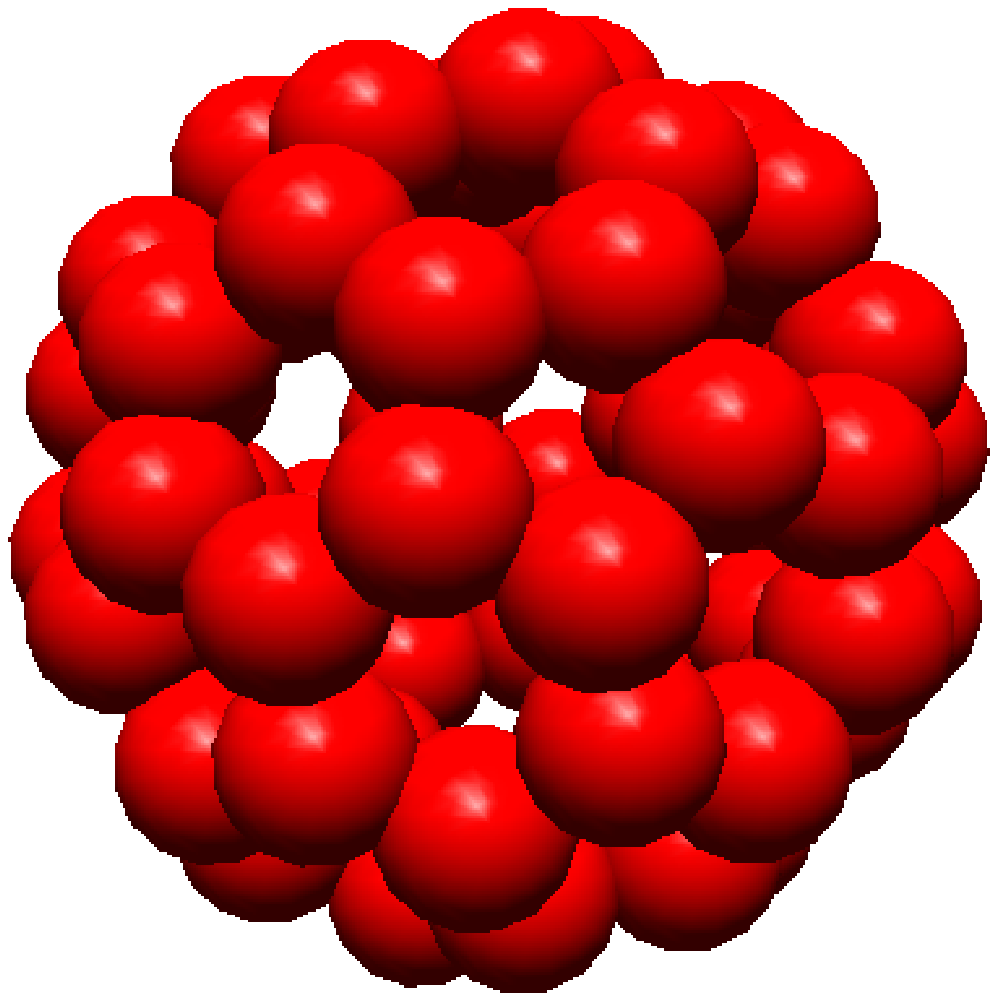}}}
\caption{ (a) The structure of the fullerene (C$_{60}$) which served as a model of a hollow superheavy nucleus with
$Z=120$ as proposed by Walter Greiner. The edges between two adjacent hexagons are double bonding.
(b) Following the analogy devised by Greiner, the Carbon atoms of the buckminsterfullerene are replaced by $\alpha$
clusters.}
\label{figfullerena}
\end{figure}

\begin{figure}[t]
\centerline{\includegraphics[width=7.75cm]{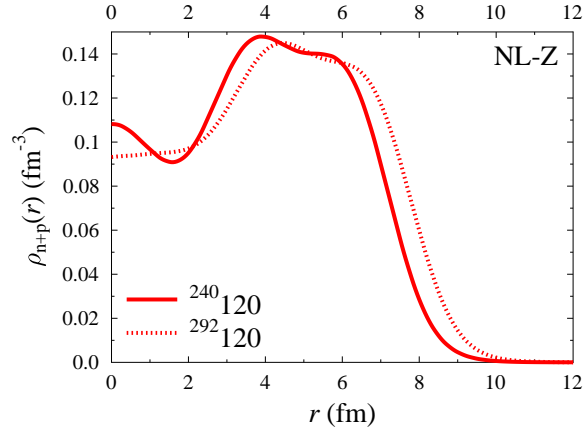}}
\caption{ Total density profiles of the superheavy isotopes of $Z$=120 : $N$=120 (solid line) and $N$=172 (dashed line).
Calculations are made within the spherical RMF model with the NLZ parametrization \cite{reinh86}.}
\label{rmf120}
\end{figure}

The $\alpha$-fullerene is a nuclear aggregate, hollow inside, with $60$ $^4$He nuclei
distributed on the vertices of 20 hexagons and 12 pentagons in a manner analogous to
the  buckminsterfullerene (C$_{60}$) \cite{cns16} (see Fig.\ref{figfullerena}(a)).
Taking as an example the nucleus $Z=$120, $N=$184, Greiner suggested that apart of the
protons and neutrons distributed over 60 $\alpha$-particles, the rest of 60 neutrons
(neglecting the last 4 neutrons) are insuring the additional bonding, i.e. one neutron per alpha
\cite{gre01}. He speculated that the two bond lengths that are 
manifest  in C$_{60}$ have their nuclear counterpart in the presence of neutrons that are not
associated in $\alpha$ particles. As Greiner asserted : {\it"such a structure would immediately 
explain the semi-hollowness of that superheavy nucleus``} \cite{gre01}. An artist view of the Greiner fullerene
is given in Fig.\ref{figfullerena}(b). The RMF framework applied by Greiner 
and collab. in 2002 to the nucleus $^{292}$120 suggests the existence of a pronounced depletion of matter in the
interior of this superheavy nucleus \cite{mis02} (see Fig.\ref{rmf120}). We compare this old case with a new calculation 
for the $N=Z$ superheavy $^{240}$120 where the depletion is also visible.

The $\alpha$-particle model of nuclei is a ubiquitous presence in the literature 
since the early days of nuclear physics \cite{wef37,ros49,paul65}. In this picture nuclei are 
viewed as crystaline structures formed of closely packed structureless spherical $\alpha$-particles,
viz. a classical assumption is made on the nuclear wave function which exhibits strong four-particle
correlations (quartet correlations). Each structure is characterized by a number of bonds or pairs 
of adjacent particles. As concluded in a study on $\alp$-conjugate
$N=Z$ nuclei \cite{oertz06}, the systematics of the binding  energy is satisfactorly described for 
light nuclei in the $\alp$-particle model.

On the other hand, over the years the possible existence of bubble and semi-bubble shapes of heavy and superheavy nuclei was 
also frequently discussed in the literature \cite{wils46,siembe67,wong73,mtw97,dbgd03}. Very recently RMF calculations  \cite{AAB16}
have shown the appearence of low nucleonic density in the central regions of lighter nuclei :  $^{22}$O
and  $^{34,36}$Si were emphasized as good candidates of being spherical bubble nuclei, whereas $^{24}$Ne, $^{32}$Si and $^{34}$Ar
were classified as deformed bubble nuclei.

\section{Stability of fullerene-like nuclei}

It was argued that since the $\alp$-particle model provides an explanation of the binding energy systematics for $N=Z$ nuclei
and the quartetting properties in the ground state, the liquid-drop model qualifies for an approximate description of the 
binding energies in terms of the mass number (semi-empirical mass formula for $\alp$-nuclei) \cite{oertz06}.
The liquid drop model framework applied to nuclei with a strongly non-uniform distribution of matter inside, as it is the case 
for bubble-like nuclear shapes, requires some modifications compared to the standard case of "normal'' nuclei
with highly uniform distribution of nuclear matter. 
Below we present a rough estimation of the LDM energy for the same superheavy nucleus taking into account only the Coulomb and surface energy.

\begin{figure}[h]
\centerline{\includegraphics[width=4.5cm]{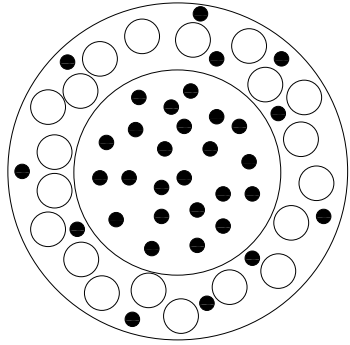}}
\caption{ Artist view of the semi-bubble nucleus resembling the endohedral cluster fullerene.
Open circles are $\alp$-particles while small black circles represent neutrons. The core region has radius $R_1$ and contains only neutrons. 
The shell extends from $R_1$ to $R_2$ and contains all protons and an equal number of neutrons bound 
in $\alp$ clusters, plus an additional number of neutrons that might contribute to the above mentioned double bondings.}
\label{artful}
\end{figure}

For a charge distribution $\rho(r)$, the Coulomb potential energy is
\begin{equation}
E_{C}=\frac{1}{2}\int dr\int dr^{\prime}\frac{\rho_{c}(\bd{r})
\rho_{c}(\bd r^{\prime})}{|\bd{r}-\bd{r^{\prime}}|}\label{ecoul}\end{equation}
where the charge density of the semi-bubble reads 
density 
\begin{equation}
\rho_{c}(\bd{r})=\rho_{c1}\Theta(R_{1}-r)
+\rho_{c2}[\Theta(R_{2}-r)-\Theta(R_{1}-r)]
\label{bubdenscoul}
\end{equation}
The charge density of the core, containing $Z_1$ protons is
$$\rho_{c1}=\frac{3}{4\pi}\frac{Z_{1}}{R_{1}^{3}}$$
and of the outer shell, containing $Z_{2}$ protons is
$$\rho_{c2}=\frac{3}{4\pi}\frac{Z_{2}e}{R_{2}^{3}-R_{1}^{3}}$$
such that
\beq
Z_1=\int_0^{R_1}\rho_{c1}d\bd{r},~~~
Z_2=\int_{R_1}^{R_2}\rho_{c2}d\bd{r}
\eeq
We use the RMF input for charge densities
from Ref.\cite{mis02}.
After some lengthy calculations we arrive at the following
expression of the Coulomb energy
\beq
E_C=\frac{3}{5}\frac{(Ze)^{2}}{R_{2}}
\frac{p^3(1-q)[p^2(3-2q)-5]+2}{2[1-p^3(1-q)]^2}
\label{coulomb}
\eeq
where by $p=R_1/R_2$ we denote the "breathing deformation`` and by $q=\rho_1/\rho_2$ the 
core density to outer shell density ratio.

Assuming a similar dependence for the total density, the surface energy can be calculated using the Yukawa-plus-exponential
interaction \cite{krappe79}, which for the particular case of a semi-bubble nucleus splits into the contribution from the
external surface, internal surface and mixed terms \cite{mismar}
\beq
E_S=E_{{\rm Y+E}~1}+E_{{\rm Y+E}~2}+E_{{\rm Y+E}~12}+E_{{\rm Y+E}~21}
\label{surface}
\eeq
The diagonal terms are casted in the form
\beqa
E_{{\rm Y+E}~i=1,2}&=&c_S(1-\kappa_SI^2)
\left (\frac{R_i}{r_0}\right )^2(1-q\delta_{i,1})^2\times\nn\\
&&\left [1-3\left(\frac{a}{R_{i}}\right )^{2}
+\left(1+\frac{R_{i}}{a}\right )\left(2+3\frac{a}{R_{i}}
+3\left(\frac{a}{R_{i}}\right )^{2}\right )e^{-2R_{i}/a}
\right ]
\eeqa
whereas for the mixed term  we get
\beqa
E_{\rm Y+E~12}&=&\frac{2c_S(1-\kappa_SI^2)}{a r_0^2}(1-q)(R_1R_2)^{3/2}\nn\\
&\times&\left [ \frac{R_1}{a}\bessi{1/2}{\frac{R_1}{a}}\bessk{3/2}{\frac{R_2}{a}}
-\frac{R_2}{a}\bessi{3/2}{\frac{R_1}{a}}\bessk{\oh}{\frac{R_2}{a}}
-3\bessi{3\over 2}{\frac{R_1}{a}}\bessk{3\over 2}{\frac{R_2}{a}}   
\right ]\nn\\
\eeqa
The last term plays a crucial role in the stability against coupled quadrupole oscillations
of the inner and outer surfaces \cite{mismar}.
The physical constants and the meaning of the special functions entering the above formulas 
are given in Ref.\cite{moel95}

We employ the above macroscopic formalism to the case illustrated in Fig.\ref{artful}. The entire charge $Z$
is assumed to be distributed in the outer layer together with an equal number of neutrons, i.e. 
$N_{\rm shell}=Z$, both species being associated in $Z/2$ $\alpha$-particles. To this shell  
$\Delta N_{\rm shell}$ neutrons are added ($N_2=Z+\Delta N_{\rm shell}$), which according to Greiner's scenario insure
the supplementary bonding between the $\alpha$-particles. 
The rest of neutrons, i.e. $N_1=N-N_{\rm shell}-\Delta N_{\rm shell}$ are distributed in the 
"hollow`` region of the semi-bubble nucleus. 

\begin{figure}[t]
\centerline{\includegraphics[width=8.cm]{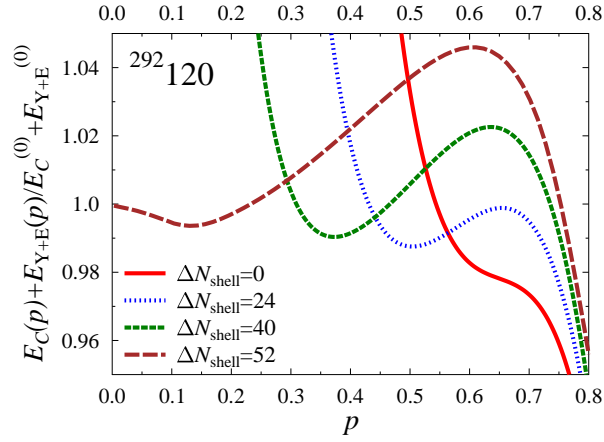}}
\caption{ 
Dependence of the total LDM energy surface (surface + Coulomb) to the equivalent spherical value on the breathing deformation
for the superheavy nucleus $^{292}$120.
Along with the 60 ''$\alp$-particles`` the outer shell is doped with $\Delta N_{\rm shell}=$ 0 (solid curve), 
24 (blue dashed), 40 (green dashed) and 52 (brown dashed) neutrons.}
\label{ldm_bubble}
\end{figure}

We plot in Fig.\ref{ldm_bubble} the sum of the Coulomb (\ref{coulomb}) and surface (\ref{surface}) energies as a function  
of the ratio $p=R_1/R_2$ relative to the equivalent spherical values  $E_{\rm S}^{(0)}+E_{C}^{(0)}$.  
For an outer shell composed of equal number of protons and neutrons ($\Delta N_{\rm shell}$=0) a 
bubble-like state with $R_1\approx 0.6 R_2$ is unstable. The addition of neutrons in the outer shell and the dilution
of the neutron fluid in the core region produces a metastable state with the breathing deformation $p$
decreasing with increasing $\Delta N_{\rm shell}$. Therefore, a perfect bubble ($N_1=N$, $N_2=0$) is the most favourable 
configuration for the nucleus $^{292}$120 in this version of the liquid drop model, 
with a completely hollow central region of small radius ($R_1 < 0.2 R_2$ ).

\section{$Q$-balls of $\alp$-clusters as nontopological solitons}

Very recently we proposed with Walter Greiner a RMF model which includes, besides the Dirac-Fermi fields describing 
standard baryonic matter composed of protons and neutrons, also $\alp$-particles described by a scalar complex field
such that the corresponding Lagrangean is allowed to contain quartic and sextic self-interaction terms \cite{misgre14,mmg17}.
The non-relativistic counterpart of these interactions is represented by two- and three-body zero-range interactions between
$\alp$-particles with strengths fixed by the $\alp-\alp$ scattering lengths extracted via the Calogero equation (see \cite{mmg17}
for more details). 
In the present work we consider the simplified version of the model (without scalar $\sigma$ and vector $\omega$ meson fields),
and take into account the electromagnetic nature of $\alp$-particles, i.e. we consider the case of the charged $Q$-ball
\cite{lee89,gulam14}. These physical objects are extensions of the simple $Q$-balls \cite{colem85,mish93b}. 
The stability of such an aggregate of $\alp$-particles results from the balance between the non-linear self-interactions of the
complex scalar field and the Coulomb repulsion between the $\alp$-particles. In our recent work \cite{mmg17} we reported some interesting conclusions
regarding the stability of neutral $Q$-balls if the meson fields are removed: 
1) the tendency for over-binding is weakened, due
to the absence of the $\sigma$-field attraction; 2) the equation-of-state (EOS) becomes much softer, due to the absence of the $\omega$-field
repulsion, making thus the $Q$-ball more easy to compress; 3) attractive quartic self-interactions, supplemented by repulsive sextic self-interactions
lead to stable configurations.

Below we consider a complex scalar field $\phi$, which describes $\alp$-particles of charge $q=+2e$  
coupled to a U(1) gauged (electromagnetic) field $A_\mu$. The Lagrangean density is 
\beq
{\cal L}=
(\partial^\mu\phi+iqA^\mu\phi)^*(\partial_{\mu}\phi+iqA_\mu\phi)
-U(|\phi|^2)-\frac{1}{4}F_{\mu\nu}F^{\mu\nu}
\label{lagralfa}
\eeq
where $F_{\mu\nu}=\partial_\mu A_\nu-\partial_\nu A_\mu$ is the electromagnetic field strength and the scalar potential function
is chosen in the form
\beq
U(|\phi|^2)={M_\alp}^2|\phi|^2+g_2|\phi|^4+\ot g_3|\phi|^6
\label{qbalscalpot}
\eeq
where $M$ is the $\alp$-particle mass and nonlinear terms describe the self-interaction of the scalar field.

Assuming for the complex scalar field a time-dependence of the form 
\beq
\phi(r,t)=\frac{1}{\sqrt{2}}\varphi(r)e^{-i{\cal\omega} t}
\eeq
and defining the radial function
\beq
\gamma(r)=\omega-q A_0(r)
\eeq
one can represent the Lagrangean density in the form 
\beq
{\cal L}=\oh\left [ \frac{1}{q^2}\left ( \frac{d\gamma}{dr}\right )^2 +\varphi^2\gamma^2
-\left ( \frac{d\varphi}{dr}\right )^2 - 2U(\varphi)\right ]
\label{actint}
\eeq
Introducing the canonical variables 
$(\phi,\pi)$ and $(A_0,{\it \Pi})$, where
\beq
\pi=\frac{\partial {\cal L}}{\partial \dot{\phi}},~~~
{\it \Pi}=\frac{\partial {\cal L}}{\partial \dot{A}_0},
\eeq
the "differential Hamilton function''\cite{Wentz49} 
\beq
{\cal H}=\pi\dot{\phi}+\pi^*\dot{\phi}^*+{\it \Pi}\dot{A}_0-{\cal L}
\eeq
is eventually casted in the form 
\beq
{\cal H}=\oh\left [ \frac{1}{q^2}\left ( \frac{d\gamma}{dr}\right )^2 +\varphi^2\gamma^2
+\left ( \frac{df}{dr}\right )^2 + 2U(\varphi)\right ]
\eeq
The $\alp$-particle density reads \cite{mmg17}
\beq
\rho=j_0=i\left ( \frac{\partial {\cal L}}{\partial({\partial_\mu\varphi^*})}\varphi^*
- \frac{\partial {\cal L}}{\partial({\partial_\mu\varphi})}\varphi\right )=g\varphi^2
\label{densqbal}
\eeq

The corresponding equations of motion are obtained by varrying the action integral \cite{Wentz49}
\beq
I=\int dt\int d\bd{r} {\cal L}(\varphi,\varphi^\prime,\gamma,\gamma^\prime)
\eeq
with respect to  $\varphi$ and $\gamma$ at fixed $\omega$, i.e.
\beq
{\varphi}{''}+\frac{2}{r}{\varphi}'+g^2\varphi-\frac{\partial U(\varphi)}{\partial\varphi}=0
\label{eqmphi}
\eeq
\beq
{\gamma}{''}+\frac{2}{r}{\gamma}'-q^2\varphi^2\gamma=0
\label{eqmg}
\eeq
The solutions of the above set of second-order non-linear differential equations were discussed in the literature 
(see for example \cite{lee89}) and concluded that for a range of values of the coupling constants ($g_2, g_3$ in
our case) the charge is pushed to the surface, the topological configuration of the gauged $Q$-ball resembling thus
a semi-bubble. In the case of a neutral $Q$-ball ($q=0$), the complex scalar field is constant inside the ball, i.e. 
$\varphi=\varphi_0$.
We take for the squared charge, $q^2=4\times1.44/(\hbar c)=0.029$

In Ref.\cite{lee89} it was established that for a  $Q$-ball with total number of $\alp$-particles $Q$ 
the radius can be expressed in  the compact form
\beq
R=\left [ \frac{3Q}{4\pi\varphi_0\sqrt{2U(\varphi_0})}\right ]^{1/3}\left ( 1+\frac{C^{2/3}}{45}\right );~~
C={3e^3Q}\sqrt{\frac{2\pi\varphi_0^4}{U(\varphi_0)}}
\label{radbal}
\eeq

\begin{figure}[t]
\centerline{\includegraphics[width=7.cm]{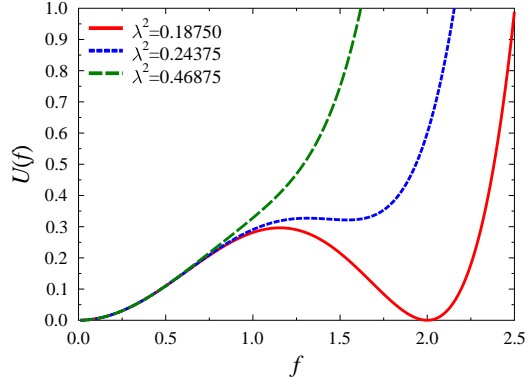}}
\caption{Scalar potential $U(f)$ for three different values of $\lambda^2$.}
\label{scalar}
\end{figure}

Since the strengths $g_2$ and $g_3$ assume large values we rescale the variables and the physical constants 
according to
\beq
\varphi=\frac{M_\alpha}{\sqrt{|g_2|}}f,~~\gamma=M_\alpha g,~~~q^2=|g_2|e^2
\eeq
such that the scalar potential (\ref{qbalscalpot}) assume the simple form used in the literature \cite{lee89,gulam14}
\beq
U(f)=\oh f^2-\frac{1}{4}f^4+\frac{\lam^2}{6}f^6
\label{scalpotlws}
\eeq
where the strength $\lambda$ is related to $g_2$ and $g_3$ via
\beq
\lam^2=\frac{g_3M_\alpha^2}{4g_2^2}
\eeq
As argued by Coleman \cite{colem85} the scalar potential fulfills $U(f)\geq 0$, a condition 
which sets an inferior bound on the 
dimensionless constant $\lambda$, i.e. $\lam^2>3/16$. 
In Fig.\ref{scalar} we displayed the scalar potential (\ref{scalpotlws}) first for this limiting value (red curve).
We observe that in this case $U(f)$ has two degenerate minima at $f=0$ and at $f\approx 2$. By increasing  $\lam^2$,
this second, non-trivial minimum, becomes shallower (blue curve) and eventually dissapears (green curve). For the last case,
when $\lam^2\approx 0.47$, the corresponding cuartic $g_2$ and sextic $g_3$ strengths are in the range of values established 
by the elastic scattering phase-shift analysis carried out in our previous and last work in collaboration with Walter Greiner \cite{mmg17}.
More precisely the term that mimicks the two-body interaction is attractive and the one corresponding to the three-body
interactions is repulsive with a value very close to the one used by us in the paper on neutral $Q$-balls.

We solved the system (\ref{eqmphi}-\ref{eqmg}) in the new representation with  dimensionless quantities :
\beqa
&&{f}{''}+\frac{2}{r}{f}'+(g^2-1)f-g_2f^3-\lam^2f^5=0\nn\\
&&{g}{''}+\frac{2}{r}{g}'-e^2f^2g=0
\label{eqmphig}
\eeqa

The system (\ref{eqmphig}) was treated as a non-linear coupled-channel problem with interior and exterior boundary conditions. 
In order to obtain a reasonable guess for $f(0)$ we make use of the above mentioned  positiveness condition on $U(f)$ for all $r$.
In our numerical experiments we take $f(0)<\sqrt{3}/2\lam$. We obtained a guess for $g(0)$ by solving the second equation of
the system (\ref{eqmphig}) in the thin-wall approximation and applied the l'Hopital rule for $r\rightarrow 0$. 
In order to find finite-energy solutions of the above equations of motion, the following initial values are assumed for the 
derivatives
\beq
f'(0)=g'(0)=0
\eeq
The exterior boundary condition was specified at $r=R$ where we assume that the nonlinearities cease to be important. 
The radius $R$ was taken to be the one provided by the estimation given in eq.(\ref{radbal}). Consequently the two equations
from (\ref{eqmphig}) are amenable to analytic solutions:
\beqa
f(r) &\sim& e^{-kr}U(1+\kappa,2,2kr),~~~\omega<M_\alpha\\
f(r)  &\sim& \frac{K_1(\sqrt{\beta r})}{\sqrt{r}},~~~\omega=M_\alpha
\eeqa
\beq
g(r)=\omega-\frac{e^2 Q}{r}
\eeq
Above, $U$ is the confluent hypergeometric function, $K_1$  the first-order modified Bessel function \cite{AS64} and 
\beq
k=\sqrt{M_\alpha^2-\omega^2},~~~\kappa=\frac{\omega e^2 Q}{kr},~~~\beta=8e^2QM_\alpha
\eeq
The numerical implementation consists in integrating a system of four ordinary differential equations, as results 
from (\ref{eqmphig}), with the help of the Runge-Kutta method of order 5 and 6 \cite{butch08}.

\begin{figure}[t]
\centerline{\includegraphics[width=8.cm]{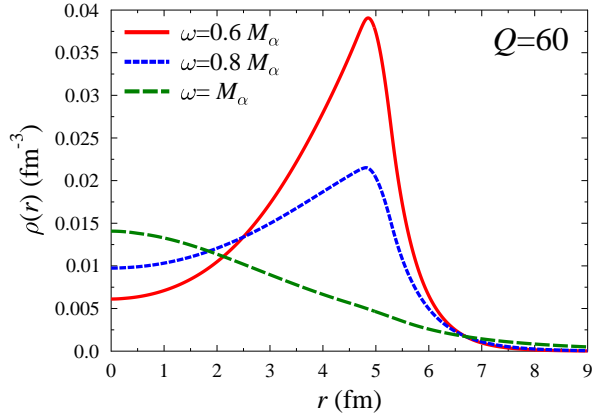}}
\caption{Profile of the $\alpha$-particle condensate with $Q$=60 for three values of the frequency $\omega$.
We set $\lam^2$ to the value corresponding to the green curve in Fig.\ref{scalar}. }
\label{profile}
\end{figure}

The profile (\ref{densqbal}) is ploted in Fig.\ref{profile} for values up to the maximum $\omega=M_\alpha$.
We choosed the following values for the cuartic and sextic strengths: $g_2=-30000$, $g_3=4786193.6$ fm$^2$.
For large $\omega$ ($\omega\rightarrow M_\alpha$) the bubble dissapears. Thus, a bubble structure 
of the $\alpha$ condensate is favoured by small $\omega$. According to Ref. \cite{kus97} this circumstance happens 
for large $Q$ when the energy of the non-topological soliton realize a minimum in the $\omega$ coordinate. 
However, it is the repulsive sextic interaction that it is mainly responsible for the depletion of the scalar field inside
the $Q$-ball.

\section{Summary and outlook}

It is natural to ask how can one validate such an atypic nuclear shape. Due to the 
hollow spherical shape and of the large number of $\alp$ particles this aggregate might 
easily develope particular rotational and vibrational modes of excitation with a different 
manifestation compared to normal nuclei. The investigation 
of such collective modes was carried out for the C$_{60}$ fullerene within a classical
model of elastic continua for a homogenous, spherical shell of atoms of zero thickness
\cite{apo95}. A similar approach can be easily extended to $\alpha$-like  fullerenes with 
non-vanishing thickness once the elastic constants of nuclear matter are derived from an energy density functional.       
Such exotic structures in nuclei could be also discriminated by considering high-lying collective excitations :
giant resonancs are expected to manifest themselves in a different manner in fullerene-like nuclei compared to normal
fluid nuclei. One of us (\c S. M.) and Walter Greiner \cite{mis02} discussed this circumstance more than 
15 years ago. If one takes as an example the giant quadrupole resonance, we expect that the isovector axial symmetric 
mode  at its amplitude produces a further depletion of proton matter inside the nucleus
and a corresponding increase at the polar tips (see Ref. \cite{EG87}, Ch.15, Fig.2). Consequently, if $\alp$ particles
are already present at the nuclear periphery and the oscillations push them in regions of low Coulomb barriers 
i.e. at the two poles of the dynamically prolate deformed nucleus, the $\alp$-decay of the superheavy nucleus
is a highly probable  doorway channel. The decay process could take place by multiple emission of $\alp$ particles!
If on the other hand we consider the excitation of the giant dipole resonance, one should take into account that the
role of different terms in the energy density in building such collective modes in bubble nuclei can be fundamentally
different compared to normal nuclei. Indeed, if the pure neutron core of low density oscillates in anti-phase to the higher density 
outer-shell, the elastic restoring constants of the hydrodynamical Steinwedel-Jensen model \cite{EG87}
will receive a consistent contribution from the compression energy.

In this paper we discussed an ideal situation, i.e. an $N=Z$ nucleus. Since in real situations
we deal with a large excess of neutrons then, if we stick to the fullerene picture of the 
superheavy $Z$=120, one may expect that the neutrons that are not associated with the $\alpha$-particle are instead
confined inside the $\alpha$-fullerene cage. One should remind the reader that an outstanding property 
of atomic fullerenes is their ability to trap atoms, ions, clusters or small molecules \cite{cns16}.
This association of fullerenes with other species leads to the so-called endohedral cluster fullerene.
In an analogous manner we expect that the excess of neutrons wanders inside the $\alpha$ cage
instead of sharing the same peripheral region with the $\alpha$-clusters. One can push the analogy even 
further and imagine a core-like structure (with proton and neutron numbers close to the magic numbers) 
in the center of the $\alpha$-fullerene  and a gas o neutrons filling the space between the core 
and the outer shell.

Although the possibility of a fullerene-type structure consisting of $\alp$ clusters is highly speculative, 
as stated in a very recent review dedicated to the state-of-art on element $Z=$120 \cite{Hoff16}, "{\em the highly
advanced experimental technology should be used also for some experiments to search for such really exotic phenomena
in the region of SHN and beyond, which is accesible using the heaviest beams and targets}``

\section{Acknowledgments}

One of the authors (\c S.M.) is gratefull to Prof. P.-G. Reinhard for elucidating him some aspects related
to the numerical implementation of the relativistic mean-field model and to Mrs. C. Matei for assistance
with the artwork. 
\c S. M. acknowledges the financial support received from the Institute of Atomic Physics-IFA,
through the national programme PN III 5/5.1/ELI-RO, Project 04-ELI/2016 ("QLASNUC'') and the
Ministry of Research and Innovation of Romania, through the Project PN 16 42 01 05/2016.
I.N. Mishustin acknowledges financial support from the Helmholz International Center for FAIR (Germany).

\section{Bibliography}\label{secbib}\index{bibliography}
%\begin{verbatim}

\end{document}